\documentclass[
aps,pre,twocolumn,nofootinbib,superscriptaddress
]{revtex4-2}

\usepackage[T1]{fontenc}
\usepackage{amsmath,amssymb,bm}
\usepackage{graphicx}
\usepackage{xcolor}
\usepackage[hidelinks]{hyperref}
\usepackage{placeins}

\newcommand{\dd}{\mathrm{d}}
\newcommand{\abs}[1]{\left|#1\right|}

\newcommand{\LerchPhi}{\Phi_{\mathrm L}}

\begin{document}

\title{Dynamics of Flat-Top--Bubble Vector Solitons}

\author{M. O. D. Alotaibi}
\email[Contact author: ]{majed.alotaibi@ku.edu.kw}
\affiliation{Department of Physics, College of Science, Kuwait University,
Sabah Al Salem University City, P.O. Box 5969, Safat 13060,
Shadadiya, Kuwait}

\author{L. Al Sakkaf}
\affiliation{Department of Applied Sciences, College of Engineering,
Abu Dhabi University, Al-Ain 59911, United Arab Emirates}

\author{U. Al Khawaja}
\affiliation{Department of Physics, School of Science,
The University of Jordan, Amman 11942, Jordan}
\affiliation{Department of Physics, United Arab Emirates University,
P.O. Box 15551, Al-Ain, United Arab Emirates}


\begin{abstract}
We investigate the dynamics of two-component flat-top--bubble vector solitons using a variational approximation and numerical simulations. The system is described by coupled cubic-quintic nonlinear Schr\"odinger equations with repulsive intercomponent coupling. We introduce a self-induced mechanism in which a localized flat-top soliton (FTS) generates a density bubble in a second component: the localized density acts as a repulsive potential for the background field, creating a composite bound state. A background-subtracted formulation and normalized super-Gaussian profiles yield collective-coordinate equations for the component centers and an effective interaction potential for their relative separation. The potential forms a binding well that is nearly triangular over its global displacement range but smooth and locally parabolic near its minimum. The variational approximation consequently predicts harmonic internal oscillations whose frequency is independent of amplitude in the small-oscillation regime. Real-time simulations verify this weak-kick behavior with good agreement between the variational and numerical frequencies. For strong phase imprints, equating the exact kick energy to the variational binding depth predicts a kick scale \(K_{\mathrm{escape}}\) for the onset of partial FTS escape. Real-time simulations support this prediction: below this scale, the composite remains nearly intact, whereas above it, a substantial fraction of the localized component leaves the bubble.
\end{abstract}

\maketitle

\section{Introduction}
Multicomponent nonlinear-wave systems support vector solitons in which two or more field components remain bound through their mutual interactions. Coupled nonlinear Schr\"odinger equations provide a framework for such states, beginning with integrable multicomponent models and their realization in nonlinear optical media
\cite{Manakov1974,ChristodoulidesJoseph1988}.
Unlike their scalar counterparts, vector solitons may possess internal degrees of freedom associated with the relative motion of their components. Dark-soliton physics in nonlinear optics and atomic Bose-Einstein condensates has been reviewed extensively
\cite{KivsharLutherDavies1998,Frantzeskakis2010,KevrekidisBook}.
A particularly well-known composite excitation is the dark-bright soliton in a repulsively interacting binary condensate, where the density notch of the dark component creates an effective trapping region for a localized bright component. Such states have been studied theoretically and observed experimentally
\cite{BuschAnglin2001,Becker2008}.
In addition to center-of-mass motion, their components may undergo an internal relative oscillation that provides a direct dynamical signature of the binding between them
\cite{AlotaibiCarr2017}.

Flat-top solitons constitute another important class of localized nonlinear states. In cubic-quintic models, they arise from the competition between focusing cubic and defocusing quintic nonlinearities and are characterized by a broad, nearly uniform central region connected to the zero background through relatively sharp edges
\cite{Malomed2022,BirnbaumMalomed2008,AlotaibiPRE2026}.
Their extended plateaus distinguish their interaction properties from those of conventional bell-shaped bright solitons. Coupling a flat-top soliton to a field with a nonzero background therefore raises the possibility of forming an asymmetric vector state composed of a localized flat-top density and an extended background carrying a localized density depletion.

Density depletions commonly referred to as bubbles occur in several
nonlinear-wave settings. Beyond-mean-field Lee--Huang--Yang corrections
can stabilize self-bound quantum droplets
\cite{Petrov2015}
and, in related one-dimensional models, can also support bubble solutions
that appear as wide density depressions on a nonzero background
\cite{Katsimiga2023}.
The stability and dynamics of these bubbles, however, depend strongly on
the underlying model. Broad flat-top droplet states supported by the same
beyond-mean-field physics also exhibit collective modes
\cite{Tylutki2020}.
In other settings, bubbles may be supported by a localized repulsive barrier
\cite{ZengFractional2021}
or stabilized by a spatial modulation of the self-defocusing nonlinearity
\cite{ZengFlatFloor2021}.
In immiscible binary condensates, bubble-droplet states can be constructed
from pairs of domain walls, with one component filling a finite domain
depleted in the other
\cite{FilatrellaMalomedSalerno2014}.
Traveling density depressions and vortex pairs have also been studied inside
two-dimensional quantum droplets, together with their behavior upon reaching
the droplet boundary
\cite{Paredes2025,ParedesBoundary2025}.
These examples show that the mechanism responsible for creating and
supporting a bubble is an essential part of its physical characterization.

Here we introduce a related but distinct self-induced mechanism in which a localized flat-top soliton self-consistently generates a bubble in a second component. One component approaches a finite background and develops a localized density depletion, while the other forms a cubic-quintic flat-top soliton on a zero background. Through the repulsive intercomponent coupling, the localized flat-top density expels the background field from the region of overlap and thereby creates the bubble. Conversely, the depleted region acts as an effective trapping well for the localized component, allowing the two structures to form a composite bound state. The resulting object is therefore neither an externally supported bubble nor a scalar density-depleted mode, but a self-induced flat-top--bubble vector soliton.

The governing coupled cubic-quintic nonlinear Schr\"odinger equations are symmetric under interchange of the two components, whereas the vector state is asymmetric because of its boundary conditions: one component approaches a modulationally stable nonzero background, while the other vanishes at infinity. The presence of the nonzero background requires particular care when defining finite energies and applying variational methods. We remove the trivial phase factors associated with the background and formulate the system in a background-subtracted form. In analogy with renormalized Lagrangian treatments of dark solitons
\cite{KivsharKrolikowski1995},
this formulation produces finite relative energy and Lagrangian densities. It also makes the binding mechanism transparent: although the original intercomponent interaction is repulsive, the relative interaction energy is reduced when the localized flat-top density overlaps the density deficit of the bubble.

To describe the relative dynamics, we introduce normalized super-Gaussian profiles for the bubble deficit and the localized flat-top density. The displaced overlap of these profiles determines an effective interaction potential for the separation between the components. This potential forms a binding well whose minimum corresponds to the aligned configuration. Although the well is nearly triangular over much of its global displacement range, its minimum is smooth and locally parabolic. The curvature at the minimum therefore predicts a harmonic internal translational mode and, to leading order, an oscillation frequency that is independent of amplitude in the small-displacement regime.

We combine the variational description with imaginary- and real-time numerical calculations. The stationary calculations produce a localized flat-top component bound to a strongly depleted bubble, while a calculation in the decoupled case confirms that the depletion is generated by the intercomponent interaction. A weak phase gradient applied to the localized component excites the relative translational mode without initially displacing either density profile. Real-time simulations show that the components remain bound and execute internal oscillations. A frequency scan confirms the predicted amplitude independence in the small-oscillation regime, and the variational and numerical frequencies are found to be in good agreement.

For stronger phase imprints, comparing the exact kick energy with the
variational binding energy predicts a scale for partial FTS
escape. Large-domain real-time simulations support this prediction, showing
a clear separation between kicks below this scale, for which the composite
remains nearly intact, and kicks above it, for which a substantial part of
the FTS leaves the bubble.

The paper is organized as follows. Section~II introduces the coupled cubic-quintic model and its background-subtracted formulation, including the corresponding relative Lagrangian density. Section~III develops the normalized collective-coordinate variational approximation and derives the effective interaction potential and the internal-mode frequency. Section~IV presents the numerical stationary states, real-time stability tests, and phase-imprinted internal oscillations. Section~V uses the variational binding energy to predict when part of the FTS escapes under a strong kick and compares that prediction with time-dependent simulations. Section~VI summarizes the main conclusions.

\section{Background-subtracted model}

Related coupled cubic-quintic systems have been used to
describe domain walls and bubble--droplet states in immiscible binary
Bose gases \cite{FilatrellaMalomedSalerno2014}. Here we consider the
symmetric coupled cubic-quintic system

\begin{equation}
i \frac{\partial \psi_1}{\partial t}
=
-\frac{1}{2}\frac{\partial^2\psi_1}{\partial x^2}
-g\abs{\psi_1}^2\psi_1
+q\abs{\psi_1}^4\psi_1
+\kappa\abs{\psi_2}^2\psi_1,
\label{eq:original1}
\end{equation}
\begin{equation}
i \frac{\partial \psi_2}{\partial t}
=
-\frac{1}{2}\frac{\partial^2\psi_2}{\partial x^2}
-g\abs{\psi_2}^2\psi_2
+q\abs{\psi_2}^4\psi_2
+\kappa\abs{\psi_1}^2\psi_2,
\label{eq:original2}
\end{equation}
where \(g>0\), \(q>0\), and \(\kappa>0\). The cubic and quintic terms are self-focusing and self-defocusing, respectively, while the intercomponent interaction is repulsive.

Although Eqs.~\eqref{eq:original1} and \eqref{eq:original2} are symmetric
under interchange of the two components, the state considered here
satisfies
$
\abs{\psi_1}^2\rightarrow\rho_0,
\abs{\psi_2}^2\rightarrow0,
\text{as }\abs{x}\rightarrow\infty,
$
where \(\rho_0\) is the background density of the first
component. The constant phase factors generated by the background are removed through
\begin{equation}
\begin{aligned}
\psi_1(x,t)
&=
u(x,t)e^{-i(-g\rho_0+q\rho_0^2)t},
\\
\psi_2(x,t)
&=
v(x,t)e^{-i\kappa\rho_0t}.
\end{aligned}
\label{eq:gauge}
\end{equation}
Here, \(u\) is the bubble component and \(v\) is the FTS component. The resulting background-subtracted equations are
\begin{equation}
i u_t
=
-\frac12u_{xx}
-g\left(\abs{u}^2-\rho_0\right)u
+q\left(\abs{u}^4-\rho_0^2\right)u
+\kappa\abs{v}^2u,
\label{eq:sub_explicit1}
\end{equation}
\begin{equation}
i v_t
=
-\frac12v_{xx}
-g\abs{v}^2v
+q\abs{v}^4v
+\kappa\left(\abs{u}^2-\rho_0\right)v.
\label{eq:sub_explicit2}
\end{equation}
These equations preserve the density dynamics of the original system, while
$
\abs{u}^2\rightarrow\rho_0,
\abs{v}^2\rightarrow0,
\abs{x}\rightarrow\infty.
$
Since \(\rho_0-\abs{u}^2>0\) inside the bubble, the term \(\kappa(\abs{u}^2-\rho_0)v\) acts as an effective attractive potential for the localized flat-top component in the depleted region. The nonzero background must be modulationally stable. As shown in
Appendix~\ref{background_stability}, this requires
$
-g+2q\rho_0>0.
$


\section{Variational approximation}
\label{sec:VA}

For the collective-coordinate reduction, we use the renormalized Lagrangian density
\begin{align}
\mathcal{L}
={}&
\frac{i}{2}
\left(u^*u_t-u u_t^*\right)
\left(1-\frac{\rho_0}{\abs{u}^2}\right)
+
\frac{i}{2}
\left(v^*v_t-v v_t^*\right)
\nonumber\\
&-
\frac12\abs{u_x}^2
-
\frac12\abs{v_x}^2
-
\frac12
\left(-g+2q\rho_0\right)
\left(\abs{u}^2-\rho_0\right)^2
\nonumber\\
&-
\frac{q}{3}
\left(\abs{u}^2-\rho_0\right)^3
+
\frac{g}{2}\abs{v}^4
-
\frac{q}{3}\abs{v}^6
\nonumber\\
&-
\kappa\abs{v}^2
\left(\abs{u}^2-\rho_0\right)
+
\frac{\rho_0}{2}c_u^2(t).
\label{eq:VA_Ldensity}
\end{align}
Variation of the field-dependent part of Eq.~\eqref{eq:VA_Ldensity} with respect to \(u^*\) and \(v^*\) yields Eqs.~\eqref{eq:sub_explicit1} and \eqref{eq:sub_explicit2}. The final term is independent of the fields and does not modify their equations of motion. It subtracts the kinetic-energy density of the uniform background generated by the spatial phase gradient \(c_u(t)\) of the bubble ansatz.

\subsection{Normalized collective-coordinate ansatz}

We use the following super-Gaussian ansatz:
\begin{align}
u(x,t)
&=
\left\{
\rho_0
-
A_u^2
\exp\left[
-\left|\frac{x-x_u(t)}{a_u}\right|^{2p}
\right]
\right\}^{1/2}
e^{i\Theta_u(x,t)},
\label{eq:VA_u_ansatz_amplitude}
\\
v(x,t)
&=
A_v
\exp\left[
-\frac12
\left|\frac{x-x_v(t)}{a_v}\right|^{2m}
\right]
e^{i\Theta_v(x,t)},
\label{eq:VA_v_ansatz_amplitude}
\end{align}
where
\begin{equation}
\Theta_j(x,t)
=
\phi_j(t)+c_j(t)\bigl[x-x_j(t)\bigr],
j=u,v.
\label{eq:VA_phases}
\end{equation}
Here \(x_u\) and \(x_v\) are the bubble and flat-top centers, respectively; \(c_u\) and \(c_v\) are the phase gradients; and \(\phi_u\) and \(\phi_v\) are global phases. The parameters \(a_u\) and \(a_v\) are the widths. The exponents \(p\) and \(m\) determine the sharpness of the bubble and flat-top edges, respectively. In the dynamical reduction, \(x_u,x_v,c_u,c_v,\phi_u,\phi_v\)
are the time-dependent collective coordinates, whereas
\(a_u,a_v,p,m\) are fixed stationary-shape parameters. The amplitudes
\(A_u\) and \(A_v\) are determined below by normalization in terms of
the conserved norms \(N_1\) and \(N_2\). For convenience, we introduce the gamma-function factors
\begin{equation}
G_m=\Gamma\left(1+\frac{1}{2m}\right),
G_p=\Gamma\left(1+\frac{1}{2p}\right).
\label{eq:VA_Gmp}
\end{equation}

The bubble-deficit norm \(N_1\) and the localized flat-top norm \(N_2\)
are then given by
\begin{align}
N_1
&=
\int_{-\infty}^{\infty}
\left(\rho_0-|u|^2\right)\dd x
=
2a_uA_u^2G_p,
\label{eq:VA_N1}
\\
N_2
&=
\int_{-\infty}^{\infty}|v|^2\,\dd x
=
2a_vA_v^2G_m.
\label{eq:VA_N2}
\end{align}
Therefore,
$
A_u^2=\frac{N_1}{2a_uG_p}, A_v^2=\frac{N_2}{2a_vG_m}.
$
The normalized ansatz is
\begin{align}
u(x,t)
&=
\left\{
\rho_0
-
\frac{N_1}{2a_uG_p}
\exp\left[
-\left|\frac{x-x_u(t)}{a_u}\right|^{2p}
\right]
\right\}^{1/2}
e^{i\Theta_u(x,t)},
\label{eq:VA_u_ansatz}
\\
v(x,t)
&=
\left(\frac{N_2}{2a_vG_m}\right)^{1/2}
\exp\left[
-\frac12
\left|\frac{x-x_v(t)}{a_v}\right|^{2m}
\right]
e^{i\Theta_v(x,t)}.
\label{eq:VA_v_ansatz}
\end{align}

The parameters satisfy \(N_1,N_2,a_u,a_v,m,p>0\). Positivity of the bubble density further requires
\begin{equation}
0<z_u\equiv
\frac{N_1}{2a_u\rho_0G_p}
\leq1.
\label{eq:VA_bubble_constraint}
\end{equation}
The value \(z_u=1\) corresponds to complete depletion at the bubble center, whereas \(0<z_u<1\) corresponds to partial depletion.

\subsection{Averaged Lagrangian and equations of motion}

Substituting Eqs.~\eqref{eq:VA_u_ansatz} and \eqref{eq:VA_v_ansatz} into Eq.~\eqref{eq:VA_Ldensity} and integrating,
\(L=\int_{-\infty}^{\infty}\mathcal{L}\,\dd x\), we obtain
\begin{align}
L
={}&
N_1\left[-c_u\dot{x}_u+\dot{\phi}_u+\frac{c_u^2}{2}\right]
+
N_2\left[c_v\dot{x}_v-\dot{\phi}_v-\frac{c_v^2}{2}\right]
\nonumber\\
&+
L_{\mathrm{int}}(x_v,x_u)
+
C_L,
\label{eq:VA_L_full}
\end{align}
where
\begin{equation}
L_{\mathrm{int}}(x_v,x_u)
=
\frac{\kappa N_1N_2}{4a_ua_vG_mG_p}\,
I_{m,p}(x_v,x_u),
\label{eq:VA_Lint}
\end{equation}
and
\begin{align}
I_{m,p}(x_v,x_u)
={}&
\int_{-\infty}^{\infty}
\exp\left[
-\left|\frac{x-x_v(t)}{a_v}\right|^{2m}
\right]
\nonumber\\
&\times
\exp\left[
-\left|\frac{x-x_u(t)}{a_u}\right|^{2p}
\right]\dd x.
\label{eq:VA_Imp}
\end{align}
The constant \(C_L\), given in Appendix~\ref{constant_term}, collects the
gradient, cubic, quintic, and Lerch-transcendent terms. The shape parameters
\(a_u\), \(a_v\), \(m\), and \(p\) are held fixed. Since \(C_L\) is
independent of \(x_u\), \(x_v\), \(c_u\), and \(c_v\), it contributes to
the stationary energy but does not enter the Euler-Lagrange equations for
these translational collective coordinates. The averaged Lagrangian depends
on the global phases \(\phi_u\) and \(\phi_v\) only through their time
derivatives. The corresponding Euler-Lagrange equations therefore give
$
\dot N_1=\dot N_2=0.
$
Thus, \(N_1\) and \(N_2\) are conserved and may subsequently be treated as
constants. The remaining translational collective coordinates obey
\begin{align}
\dot{x}_u
&=
c_u,
\label{eq:VA_ode_xu}
\\
\dot{x}_v
&=
c_v,
\label{eq:VA_ode_xv}
\\
\dot{c}_u
&=
-\frac{\kappa N_2}{4a_ua_vG_mG_p}
\frac{\partial I_{m,p}}{\partial x_u},
\label{eq:VA_ode_cu}
\\
\dot{c}_v
&=
\frac{\kappa N_1}{4a_ua_vG_mG_p}
\frac{\partial I_{m,p}}{\partial x_v}.
\label{eq:VA_ode_cv}
\end{align}
Since \(I_{m,p}\) depends on the centers only through \(x_v-x_u\), the aligned state \(x_u=x_v\) is an equilibrium. Eliminating the phase gradients gives
\begin{align}
\ddot{x}_u
&=
-\frac{\kappa N_2}{4a_ua_vG_mG_p}
\frac{\partial I_{m,p}}{\partial x_u},
\label{eq:VA_ode2_xu}
\\
\ddot{x}_v
&=
\frac{\kappa N_1}{4a_ua_vG_mG_p}
\frac{\partial I_{m,p}}{\partial x_v}.
\label{eq:VA_ode2_xv}
\end{align}

We define the relative separation as
\begin{equation}
\ell(t)=x_v(t)-x_u(t).
\label{eq:VA_ell_definition}
\end{equation}
Using \(I_{m,p}(x_v,x_u)=I_{m,p}(\ell)\), the two center equations combine into
\begin{equation}
\ddot{\ell}
=
\frac{
\kappa\left(N_1-N_2\right)
}{
4a_ua_vG_mG_p
}
\frac{\dd I_{m,p}}{\dd\ell}.
\label{eq:VA_ell_general}
\end{equation}

\subsection{Overlap integral}

\subsubsection{Gaussian case: \(m=p=1\)}

For Gaussian profiles, the complete displacement-dependent overlap is
\begin{equation}
I_{1,1}(\ell)
=
\frac{
\sqrt{\pi}\,a_ua_v
}{
\sqrt{a_u^2+a_v^2}
}
\exp\left[
-\frac{\ell^2}{a_u^2+a_v^2}
\right].
\label{eq:VA_I_gaussian}
\end{equation}
Using
\(
G_m=G_p=G_1=\Gamma(3/2)=\sqrt{\pi}/2
\)
and substituting Eq.~\eqref{eq:VA_I_gaussian} into Eq.~\eqref{eq:VA_ell_general}, we obtain
\begin{equation}
\ddot{\ell}
+
\frac{
2\kappa\left(N_1-N_2\right)
}{
\sqrt{\pi}\left(a_u^2+a_v^2\right)^{3/2}
}
\ell
\exp\left[
-\frac{\ell^2}{a_u^2+a_v^2}
\right]
=
0.
\label{eq:VA_gaussian_ell}
\end{equation}

For small relative displacements about \(\ell=0\),
\begin{equation}
\ddot{\ell}
+
\frac{
2\kappa\left(N_1-N_2\right)
}{
\sqrt{\pi}\left(a_u^2+a_v^2\right)^{3/2}
}
\ell
+
\mathcal{O}\left(\ell^3\right)
=
0.
\label{eq:VA_gaussian_ell_linear}
\end{equation}
Therefore,
\begin{equation}
\omega
=
\left[
\frac{
2\kappa\left(N_1-N_2\right)
}{
\sqrt{\pi}\left(a_u^2+a_v^2\right)^{3/2}
}
\right]^{1/2}.
\label{eq:VA_gaussian_frequency}
\end{equation}
In the equal-width limit, \(a_u=a_v=a\),
\begin{equation}
\omega^2
=
\frac{\kappa\left(N_1-N_2\right)}
{\sqrt{2\pi}\,a^3}.
\label{eq:VA_equal_width_frequency}
\end{equation}
This has the same norm ordering and structure as Eq.~(17a) of Ref.~\cite{AlotaibiCarr2017}. In both systems, \(N_1\) is the deficit norm of the background component and \(N_2\) is the localized bright-component norm. Therefore, both calculations predict a real internal mode when the deficit norm exceeds the localized norm, with
\(\omega\propto\sqrt{\text{coupling}}\,(\text{width})^{-3/2}\sqrt{N_1-N_2}\).

\subsubsection{General super-Gaussian case}

For general super-Gaussian orders, the displaced overlap is
\begin{align}
I_{m,p}(\ell)
={}&
\int_{-\infty}^{\infty}
\exp\left[
-\left|\frac{y-\ell}{a_v}\right|^{2m}
\right]
\nonumber\\
&\times
\exp\left[
-\left|\frac{y}{a_u}\right|^{2p}
\right]\dd y.
\label{eq:VA_Imp_general}
\end{align}
In general, this integral does not reduce to a compact closed form and must be evaluated numerically.

For fixed profile shapes, the full background-subtracted energy is
\(E_{\mathrm{rel}}(\ell)=E_u+E_v+U_{\mathrm{bind}}(\ell)\), where
\(E_u\) and \(E_v\) are the bubble and FTS self-energies. These terms are
independent of \(\ell\). Therefore, the only displacement-dependent contribution is the negative of
the interaction term in the averaged Lagrangian. We define the binding
potential
\begin{equation}
U_{\mathrm{bind}}(\ell)
=
-L_{\mathrm{int}}(\ell)
=
-\frac{\kappa N_1N_2}{4a_ua_vG_mG_p}
I_{m,p}(\ell).
\label{eq:VA_Ubind_general}
\end{equation}
Since the overlap vanishes for infinitely separated profiles,
\(U_{\mathrm{bind}}(\ell)\rightarrow0\), while
\(E_{\mathrm{rel}}(\ell)\rightarrow E_u+E_v\), as
\(\abs{\ell}\rightarrow\infty\). The self-energies cancel when the aligned
and infinitely separated configurations are compared. The magnitude of the
binding energy is therefore
\begin{equation}
\abs{E_b}
=
\abs{U_{\mathrm{bind}}(0)-U_{\mathrm{bind}}(\infty)}
=
\frac{\kappa N_1N_2}{4a_ua_vG_mG_p}
I_{m,p}(0).
\label{eq:VA_Eb_general}
\end{equation}

Because both profiles are even functions about their respective centers,
\begin{equation}
I_{m,p}(-\ell)=I_{m,p}(\ell).
\label{eq:VA_Ieven}
\end{equation}
Its expansion about the aligned state is
\begin{equation}
I_{m,p}(\ell)
=
I_{m,p}(0)
+
C_2\ell^2
+
\mathcal{O}\left(\ell^4\right), \;C_2=\frac{1}{2}I_{m,p}''(0).
\label{eq:VA_Iexpansion_general}
\end{equation}
Differentiation under the integral, followed by integration by parts, gives
\begin{align}
C_2
={}&
-\frac{2mp}{a_v^{2m}a_u^{2p}}
\int_{-\infty}^{\infty}
\abs{y}^{2m+2p-2}
\nonumber\\
&\times
\exp\left[
-\left|\frac{y}{a_v}\right|^{2m}
-\left|\frac{y}{a_u}\right|^{2p}
\right]\dd y.
\label{eq:VA_C2_general}
\end{align}
Hence \(C_2<0\), showing that the overlap is maximal at \(\ell=0\), where the corresponding binding potential is minimal. Substituting Eq.~\eqref{eq:VA_Iexpansion_general} into Eq.~\eqref{eq:VA_Ubind_general} gives
\begin{align}
U_{\mathrm{bind}}(\ell)
={}&
U_{\mathrm{bind}}(0)
-
\frac{\kappa N_1N_2}{4a_ua_vG_mG_p}
C_2\ell^2
\nonumber\\
&+
\mathcal{O}\left(\ell^4\right).
\label{eq:VA_Ubind_quadratic}
\end{align}
Because \(C_2<0\), the coefficient of \(\ell^2\) is positive. Thus the
binding potential is parabolic to leading order near the aligned fixed
point. Equation~\eqref{eq:VA_Ubind_general} gives the potential at any
displacement by numerically evaluating the overlap integral, while
Eq.~\eqref{eq:VA_Ubind_quadratic} gives its approximation for small
displacements near the aligned state.

Independently, substituting Eq.~\eqref{eq:VA_Iexpansion_general} into Eq.~\eqref{eq:VA_ell_general} and retaining terms linear in \(\ell\) gives
\begin{equation}
\ddot{\ell}
+
\omega_{m,p}^{\,2}\ell
=
0,
\label{eq:VA_general_harmonic}
\end{equation}
where
\begin{equation}
\omega_{m,p}^{\,2}
=
-\frac{
\kappa\left(N_1-N_2\right)
}{
2a_ua_vG_mG_p
}
C_2.
\label{eq:VA_general_frequency_C2}
\end{equation}
Equivalently,
\begin{align}
\omega_{m,p}^{\,2}
={}&
\frac{
\kappa mp\left(N_1-N_2\right)
}{
a_v^{2m+1}a_u^{2p+1}G_mG_p
}
\int_{-\infty}^{\infty}
\abs{y}^{2m+2p-2}
\nonumber\\
&\times
\exp\left[
-\left|\frac{y}{a_v}\right|^{2m}
-\left|\frac{y}{a_u}\right|^{2p}
\right]\dd y.
\label{eq:VA_general_frequency}
\end{align}
Thus, for unequal super-Gaussian orders, calculating the internal
frequency requires only one numerical evaluation of the integral. 

\begin{figure}[t]
\centering
\includegraphics[width=\columnwidth]{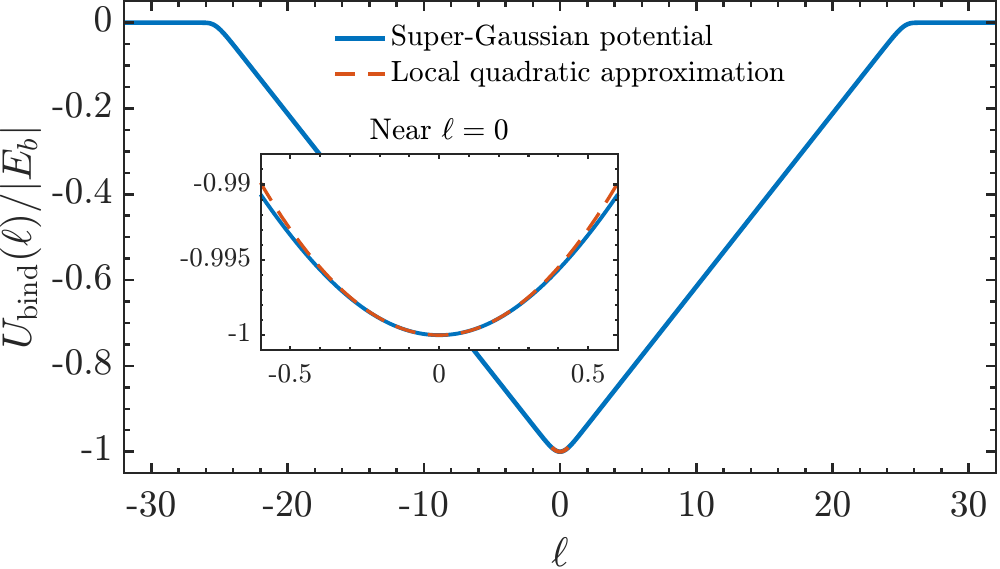}
\caption{
Normalized variational binding potential evaluated using the fitted
parameters reported in the following numerical section:
\(a_u=12.9298\), \(p=22.7550\), \(a_v=12.6487\), \(m=21.5548\),
\(N_1=30.65238\), \(N_2=30\), and \(\kappa=4\).
The solid curve is the numerically evaluated potential from
Eq.~\eqref{eq:VA_Imp_general}, and the dashed curve is the local quadratic
approximation in Eq.~\eqref{eq:VA_Ubind_quadratic}. The inset enlarges the
small-displacement region near the aligned state.
}
\label{fig:VA_binding_potential}
\end{figure}

Figure~\ref{fig:VA_binding_potential} shows the corresponding binding potential evaluated using the profile parameters obtained by fitting the super-Gaussian ansatz to the stationary numerical profiles. The fitting procedure and the resulting parameters are presented in the following numerical section; in particular, Eq.~\eqref{eq:numerical_fit_parameters} below gives \(m=21.5548\) and \(p=22.7550\), so that \(m,p\approx22\). The displaced overlap is the cross-correlation of the two super-Gaussian profiles. Because both profiles are even, this cross-correlation is equivalent to their convolution. For these large fitted orders, both super-Gaussians are nearly rectangular, and the convolution of two nearly equal rectangular pulses is approximately triangular. Since the binding potential is the negative of the overlap, the triangular overlap appears as a V-shaped potential well over the global displacement range. The finite super-Gaussian orders nevertheless round the bottom of the well. Consequently, sufficiently close to \(\ell=0\), the potential remains smooth and parabolic, as shown by the inset. This local parabolic behavior motivates the expectation of an approximately amplitude-independent internal frequency for sufficiently small oscillations.

\section{Numerical results}
\label{sec:numerics}

\subsection{Stationary flat-top--bubble state}
\label{subsec:stationary_state}

Stationary states are obtained by imaginary-time propagation of
Eqs.~\eqref{eq:sub_explicit1} and \eqref{eq:sub_explicit2} using a
split-step Fourier method with \(n_x=4096\), \(\Delta x=0.05\), and
\(\Delta\tau=5\times10^{-4}\). We use
$
g=q=4, \rho_0=1.2, \kappa=4,
$
and fix the localized norm at \(N_2=30\). The background-stability condition~\eqref{eq:mi_condition} is satisfied,
since \(-g+2q\rho_0=5.6>0\).

\begin{figure}[t]
\centering
\includegraphics[width=\columnwidth]{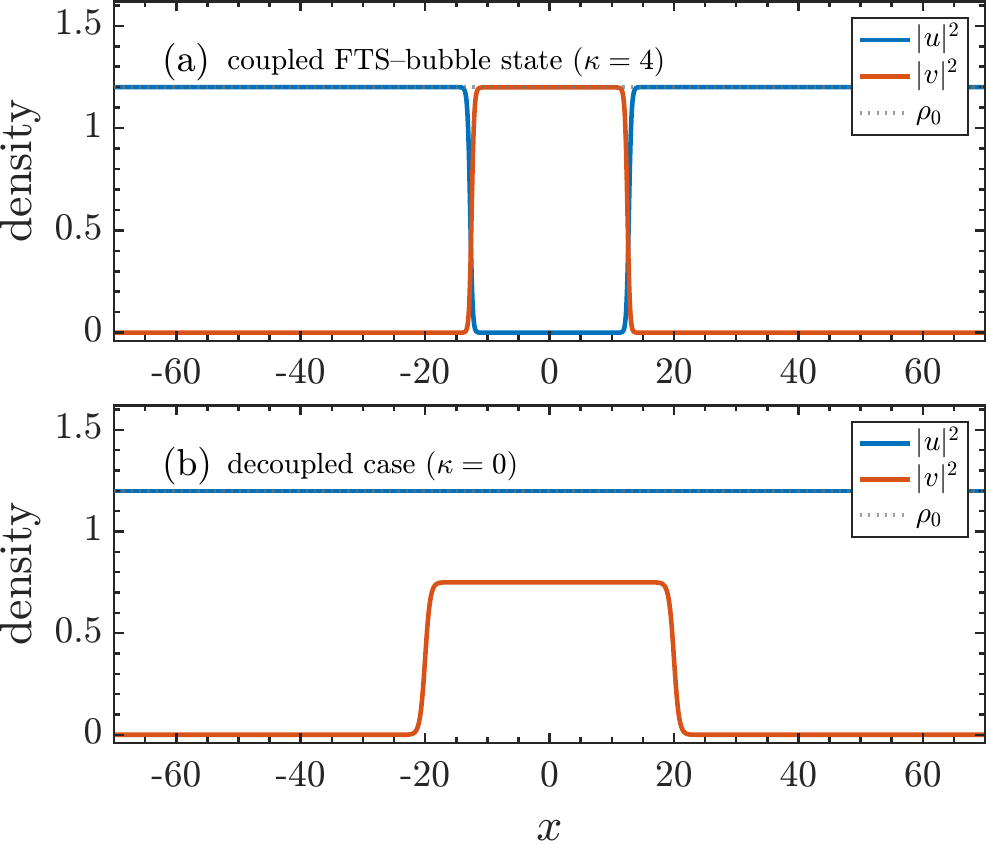}
\caption{
Stationary density profiles for \(g=q=4\), \(\rho_0=1.2\), and
\(N_2=30\). (a) Coupled flat-top--bubble state with \(\kappa=4\).
(b) Decoupled case with \(\kappa=0\). The dotted line denotes the
background density \(\rho_0\).
}
\label{fig:stationary_state_control}
\end{figure}

Figure~\ref{fig:stationary_state_control}(a) shows the converged
flat-top--bubble state. The localized component \(v\) forms a flat-top
profile and generates an almost completely depleted region in the
background component \(u\). The corresponding deficit norm is
$
N_1
=
\int_{-\infty}^{\infty}
\left(\rho_0-|u|^2\right)\dd x
=
30.65238,
$
so that \(N_1>N_2\). For comparison, Fig.~\ref{fig:stationary_state_control}(b)
shows the decoupled case, \(\kappa=0\), where \(u\) returns to the uniform
background and \(v\) becomes an isolated flat-top
soliton.

\begin{figure}[t]
\centering
\includegraphics[width=\columnwidth]{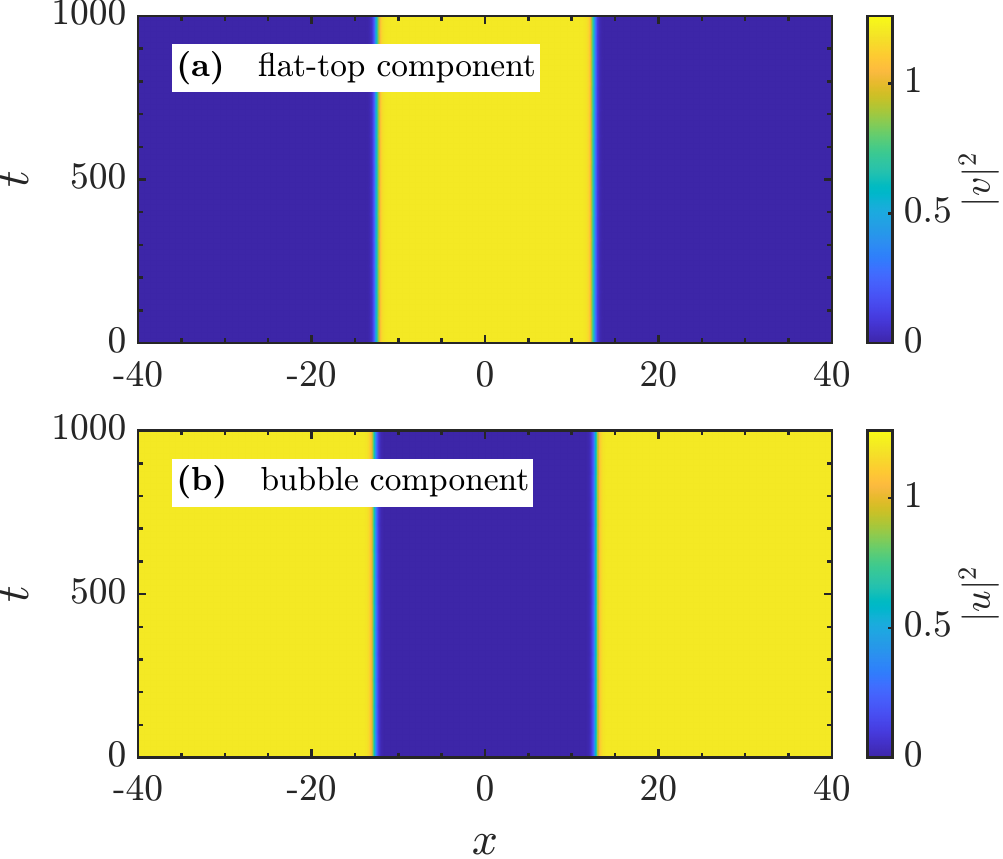}
\caption{
Stationary real-time evolution of the coupled state for
\(g=q=4\), \(\rho_0=1.2\), \(\kappa=4\), and \(N_2=30\).
(a) Density \(\abs{v}^2\) of the localized flat-top component.
(b) Density \(\abs{u}^2\) of the finite-background bubble component.}
\label{fig:RT_stationary_density_maps}
\end{figure}

We additionally test whether the converged imaginary-time state remains
stationary under the governing real-time equations. Starting directly from the
profiles in Fig.~\ref{fig:stationary_state_control}(a), without applying a
phase kick or an initial displacement, we propagate both components up to
\(T_{\max}=1000\). The space-time density maps in
Fig.~\ref{fig:RT_stationary_density_maps} show that both components remain
centered and retain their flat-top and bubble profiles throughout the
evolution.


\subsection{Internal-mode frequency}
\label{subsec:internal_frequency}

\begin{figure}[t]
\centering
\includegraphics[width=\columnwidth]{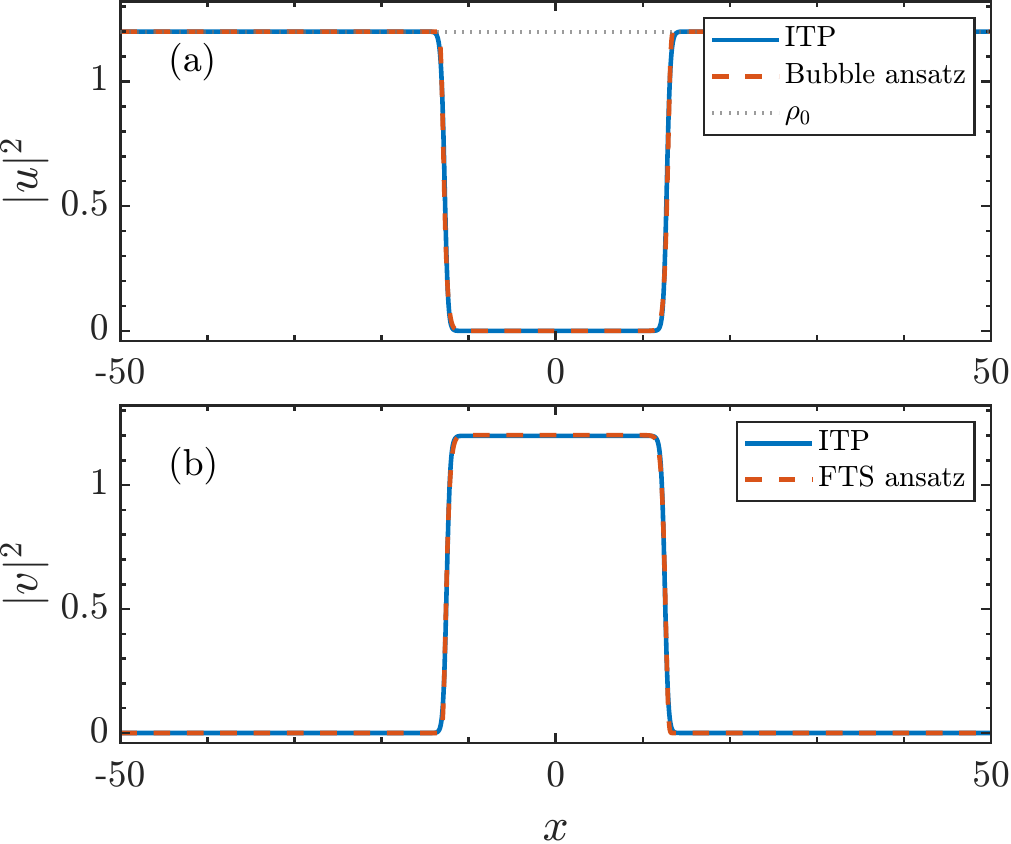}
\caption{
Comparison between the stationary profiles obtained by imaginary-time propagation (solid blue curves) and the normalized super-Gaussian ansatz (dashed orange curves) evaluated using the fitted parameters in Eq.~\eqref{eq:numerical_fit_parameters}.
(a) Background-component density \(\abs{u}^2\), showing the density bubble; the dotted line denotes the background density \(\rho_0\).
(b) Localized flat-top density \(\abs{v}^2\).
}
\label{fig:VA_numerical_fit}
\end{figure}

To evaluate Eq.~\eqref{eq:VA_general_frequency}, we first fit the stationary numerical profiles to the normalized super-Gaussian ansatz introduced in Sec.~\ref{sec:VA}. The fitted parameters are
\begin{equation}
\begin{aligned}
a_u&=12.9298, & p&=22.7550,\\
a_v&=12.6487, & m&=21.5548.
\end{aligned}
\label{eq:numerical_fit_parameters}
\end{equation}
The resulting profiles are compared with the imaginary-time propagation results in Fig.~\ref{fig:VA_numerical_fit}.

Using these fitted parameters together with \(N_1=30.65238\),
\(N_2=30\), and \(\kappa=4\) in
Eq.~\eqref{eq:VA_general_frequency} gives
\begin{equation}
\omega_{\mathrm{VA}}=0.07551,
T_{\mathrm{VA}}
=
\frac{2\pi}{\omega_{\mathrm{VA}}}
=
83.21.
\label{eq:VA_numerical_prediction}
\end{equation}

The internal mode is excited by applying a weak phase gradient to the
localized component,
\begin{equation}
v(x,0)=v_0(x)e^{iK_0x},
u(x,0)=u_0(x).
\label{eq:phase_imprint}
\end{equation}
The phase imprint changes the initial velocity of \(v\) without shifting
either density profile. For the representative case \(K_0=0.010\), the
system is evolved in real time using a split-step Fourier method, and the
relative coordinate
\begin{equation}
\ell(t)=x_v(t)-x_u(t)
\end{equation}
is monitored up to \(T_{\max}=1000\). We obtain the numerical angular frequency from the dominant positive-frequency
peak of the fast Fourier transform (FFT) power spectrum of \(\ell(t)\).
Real-time calculations with \(\Delta t=10^{-3}\) and \(5\times10^{-4}\)
give the same result to the quoted precision. We therefore find
\begin{equation}
\omega_{\mathrm{num}}
\equiv
\omega_{\mathrm{FFT}}
=
0.07948,
T_{\mathrm{num}}
=
\frac{2\pi}{\omega_{\mathrm{num}}}
=
79.06.
\label{eq:numerical_internal_frequency}
\end{equation}

\begin{figure}[t]
\centering
\includegraphics[width=\columnwidth]{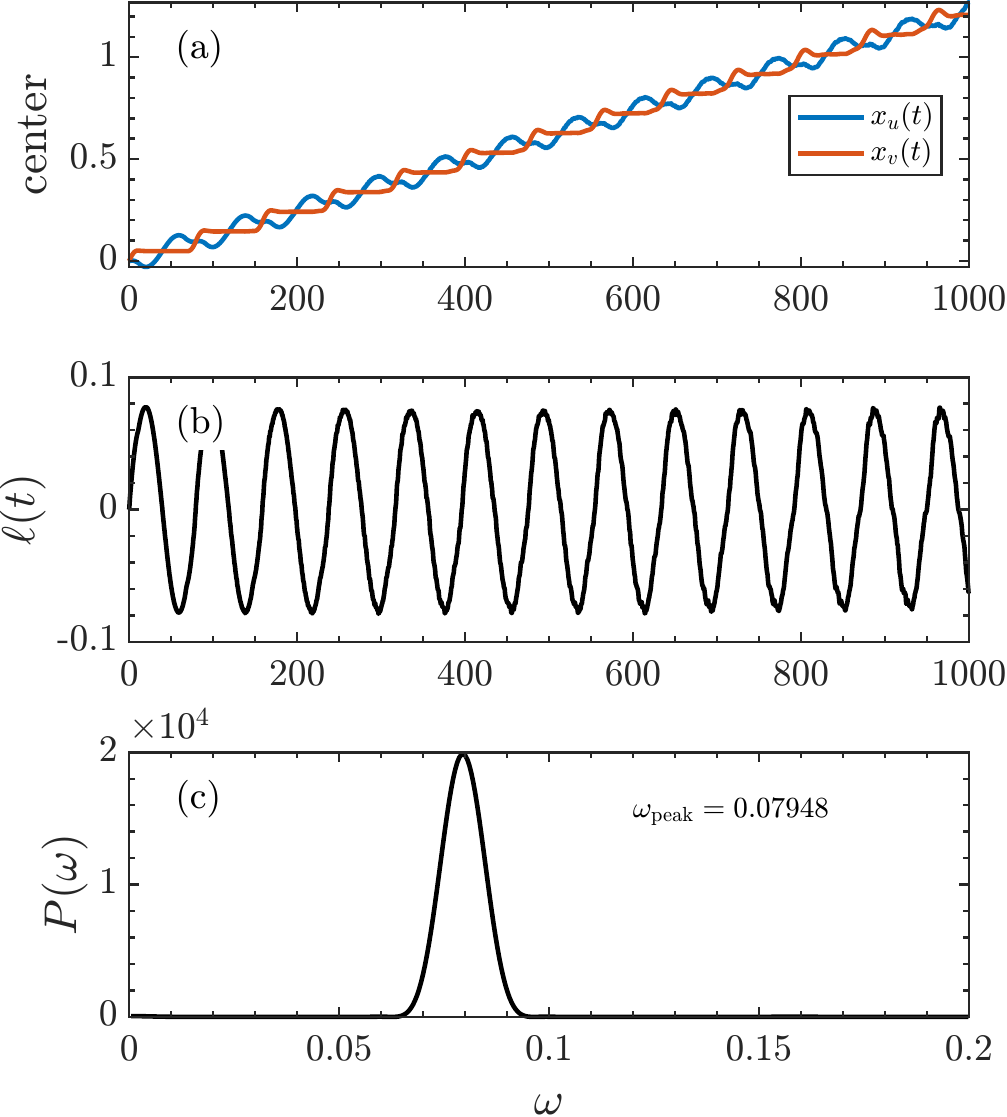}
\caption{
Phase-kicked internal motion for \(K_0=0.010\).
(a) Centers of the bubble and localized flat-top components.
(b) Relative coordinate \(\ell(t)=x_v(t)-x_u(t)\).
(c) FFT power spectrum of the relative motion. The dominant peak gives
\(\omega_{\mathrm{num}}\equiv\omega_{\mathrm{FFT}}=0.07948\).
}
\label{fig:internal_oscillation}
\end{figure}

Figure~\ref{fig:internal_oscillation} shows the component centers, their
relative motion, and the FFT peak used to obtain
Eq.~\eqref{eq:numerical_internal_frequency}.
The relative difference between the numerical and variational frequencies is
\begin{equation}
\frac{\abs{\omega_{\mathrm{num}}-\omega_{\mathrm{VA}}}}
{\omega_{\mathrm{num}}}
\times100\%
=
4.99\%.
\end{equation}

\begin{figure}[t]
\centering
\includegraphics[width=\columnwidth]{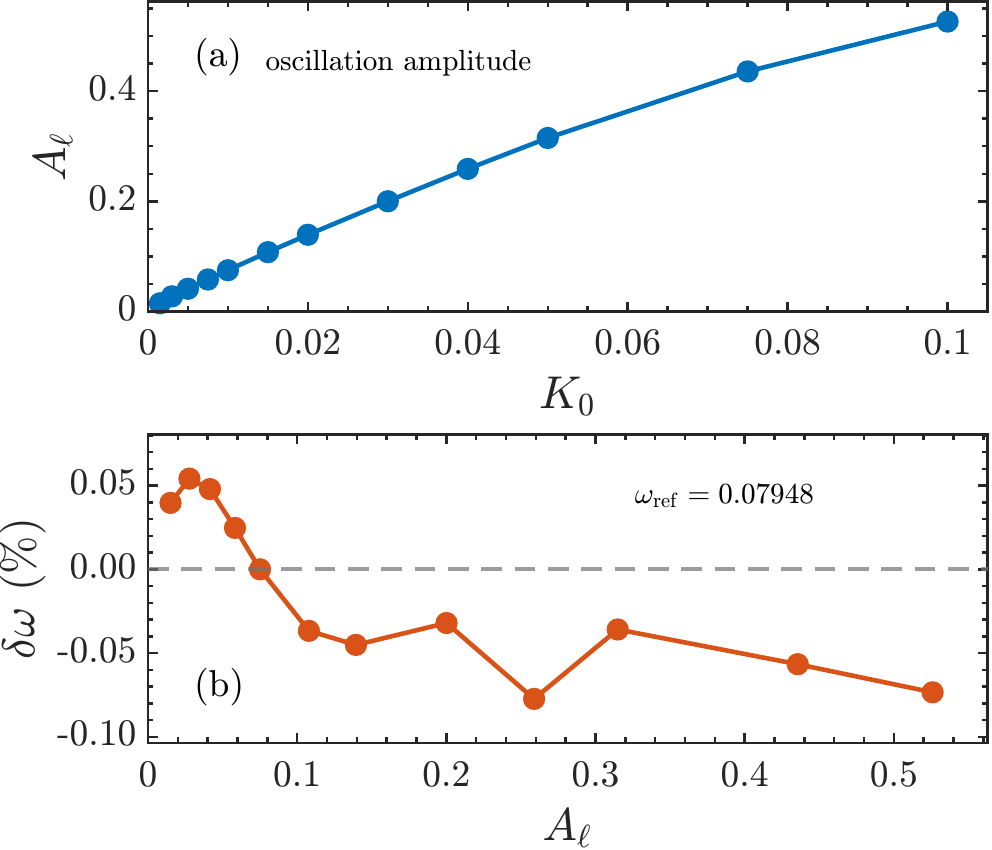}
\caption{
Phase-kick scan of the internal mode.
(a) Measured oscillation amplitude \(A_\ell\) as a function of \(K_0\).
(b) Relative FFT frequency shift
\(\delta\omega=100(\omega_{\mathrm{FFT}}-\omega_{\mathrm{ref}})/
\omega_{\mathrm{ref}}\), using \(\omega_{\mathrm{ref}}=0.07948\) from
\(K_0=0.010\). The dashed
zero line means no change from this reference. The total variation is only
\(0.13\%\), showing that the internal-mode frequency is nearly independent
of oscillation amplitude over the scanned range.
}
\label{fig:internal_frequency_scan}
\end{figure}

To test the dependence on oscillation amplitude, we repeat the calculation
for 12 phase kicks between \(K_0=0.0015\) and \(0.100\). Since the kick
sets the initial velocity rather than the displacement, each run is
characterized by the measured oscillation amplitude \(A_\ell\).
Figure~\ref{fig:internal_frequency_scan}(a) shows that \(A_\ell\) grows
smoothly with \(K_0\). Figure~\ref{fig:internal_frequency_scan}(b) shows
the relative frequency shift
\(\delta\omega=100(\omega_{\mathrm{FFT}}-\omega_{\mathrm{ref}})/
\omega_{\mathrm{ref}}\) from the representative value
\(\omega_{\mathrm{ref}}=0.07948\), obtained for \(K_0=0.010\). Across the
tested interval \(0.0151\leq A_\ell\leq0.5260\), the total frequency variation
is only \(0.13\%\). Thus, the internal-mode frequency is effectively
independent of amplitude over the tested interval.

\begin{figure}[t]
\centering
\includegraphics[width=\columnwidth]{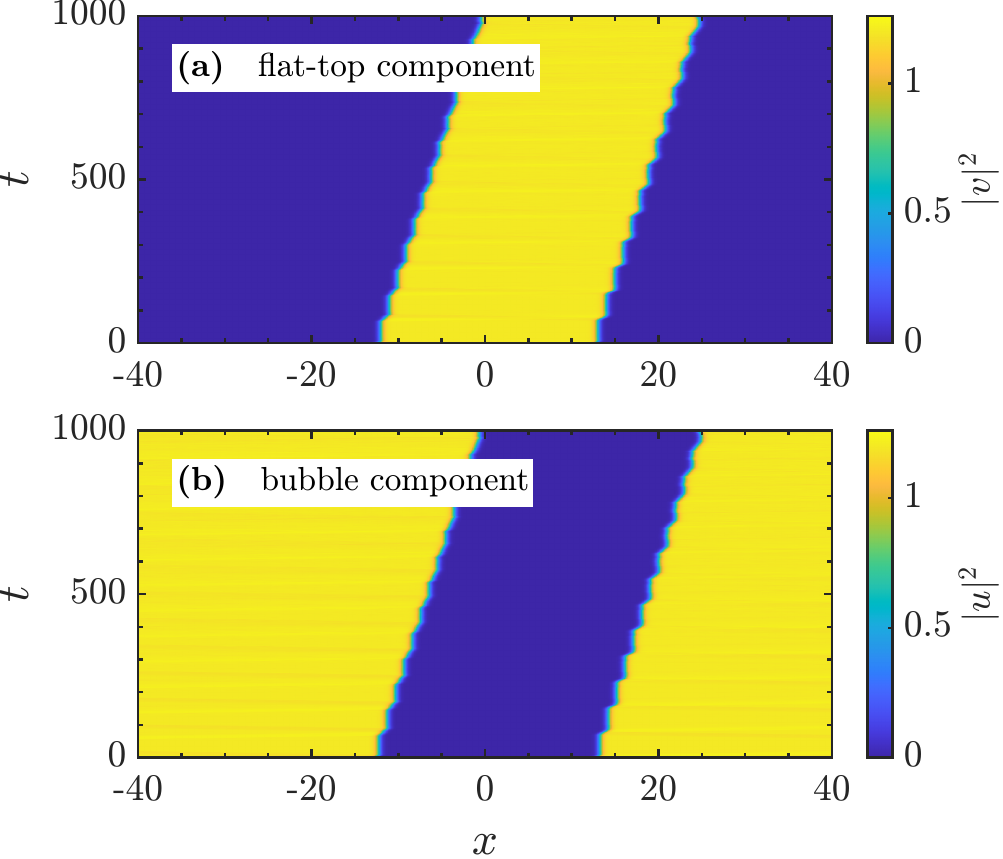}
\caption{
Real-time density maps over the displayed interval \(-40\leq x\leq40\) for
the largest phase kick in the internal-mode scan, \(K_0=0.100\).
(a) Localized flat-top component \(\abs{v}^2\).
(b) Finite-background bubble component \(\abs{u}^2\).
The phase imprint produces a common drift together with a bound relative
oscillation.
}
\label{fig:phase_kick_density_maps}
\end{figure}

For a direct space-time view of the strongest excitation in the scan,
Fig.~\ref{fig:phase_kick_density_maps} shows the real-time densities at the
largest applied phase kick, \(K_0=0.100\). Both components remain bound while
undergoing a common drift, with the internal oscillation appearing as a
small periodic relative motion of their density profiles.

As an independent check on the profile-fit parameters in
Eq.~\eqref{eq:numerical_fit_parameters}, we also perform a static energy
minimization in Appendix~\ref{app:static_minimization}. That calculation
determines the profile parameters directly from the background-subtracted
energy and gives \(\omega_{\mathrm{VA}}^{(E)}=0.07296\), with a relative
difference of \(8.20\%\) from \(\omega_{\mathrm{num}}\); see
Eq.~\eqref{eq:static_min_frequency}. Thus, both parameter estimates give
frequencies close to the numerical result, while the profile-fit parameters
provide the more accurate prediction.



\section{FTS response to strong phase kicks}
\label{sec:nonlinear_escape}

The weak phase imprints in Sec.~\ref{subsec:internal_frequency} probe the
nearly harmonic internal mode. We now use the same imprint,
Eq.~\eqref{eq:phase_imprint}, at much larger \(K_0\) to test when part of
the localized flat-top component begins to escape from the bubble
that initially binds it.

The phase imprint leaves the density profile \(\abs{v_0}^2\) unchanged, so
it changes only the spatial-gradient contribution to the energy. To isolate
the energy added by the kick, we subtract the original gradient energy
\(\frac12\int\abs{v_{0,x}}^2\dd x\), which is associated with the shape of
the stationary profile. The resulting change in kinetic energy is
\begin{align}
\Delta E_{\mathrm{kick}}
={}&\frac12\int_{-\infty}^{\infty}
\left[
\abs{\partial_x(v_0e^{iK_0x})}^2-
\abs{v_{0,x}}^2
\right]\dd x
\nonumber\\
={}&\frac{K_0^2}{2}\int_{-\infty}^{\infty}\abs{v_0}^2\dd x
+K_0\int_{-\infty}^{\infty}
\operatorname{Im}(v_0^*v_{0,x})\dd x
\nonumber\\
={}&\frac{N_2K_0^2}{2}.
\label{eq:kick_energy_escape}
\end{align}
In the second line, the original shape contribution \(\abs{v_{0,x}}^2\)
has canceled. The last equality is exact for the stationary state because
\(v_0\) is real up to a constant phase, so the remaining current term
vanishes, and its norm is \(N_2\).

The global binding depth has already been obtained from the
variational overlap in Eq.~\eqref{eq:VA_Eb_general}. Evaluating that
expression with the fitted profiles gives
\(\abs{E_b}=142.9968\). Equating this binding depth to
Eq.~\eqref{eq:kick_energy_escape} defines the FTS escape energy
scale
\begin{equation}
K_{\mathrm{escape}}
=
\sqrt{\frac{2\abs{E_b}}{N_2}}
=3.0876.
\label{eq:K_energy_escape}
\end{equation}
The independently energy-minimized profiles obtained in
Appendix~\ref{app:static_minimization} give
\(\abs{E_b}=142.8735\) and \(K_{\mathrm{escape}}=3.0862\). The close agreement shows that this predicted scale is insensitive to the two
parameter-estimation procedures. It should be interpreted as an energetic
onset scale, not as a sharp dissociation threshold, because the nonlinear
field dynamics may redistribute the injected energy into deformation,
radiation, and partial recapture.

To test this prediction, we propagate the phase-kicked states for
\(K_0=0,1,2,4,6,\) and \(8\). We use \(n_x=16384\) grid points with
\(\Delta x=0.05\), giving the large computational domain
\(L_x=n_x\Delta x=819.2\), together with the time step
\(\Delta t=10^{-3}\). This large domain suppresses recurrence of the
emitted branches over the displayed interval. We measure
the fraction of the conserved FTS norm inside a window of half-width \(R\)
centered on the continuously tracked original bubble,
\begin{equation}
F_{\mathrm{bubble}}(t)
=
\frac{1}{N_2}
\int_{x_b(t)-R}^{x_b(t)+R}\abs{v(x,t)}^2\dd x,
\label{eq:FTS_remaining_fraction}
\end{equation}
where \(x_b(t)\) is the center of the tracked density minimum. We choose
\(R=18\), which is larger than half the stationary FTS width
\(W_{v,\mathrm{FWHM}}\simeq25\) obtained in
Appendix~\ref{app:static_minimization}. This window contains the bound FTS
while excluding density that clearly separates from the bubble. Thus,
\(F_{\mathrm{bubble}}=1\) means that the FTS norm remains near the
bubble, while smaller values indicate that part of the FTS has escaped from
this region. This measure is directly analogous to the fraction of the
localized component remaining near a dark-bright soliton during breakup in
Ref.~\cite{AlotaibiCarr2017}.

\begin{figure}[t]
\centering
\includegraphics[width=\columnwidth]{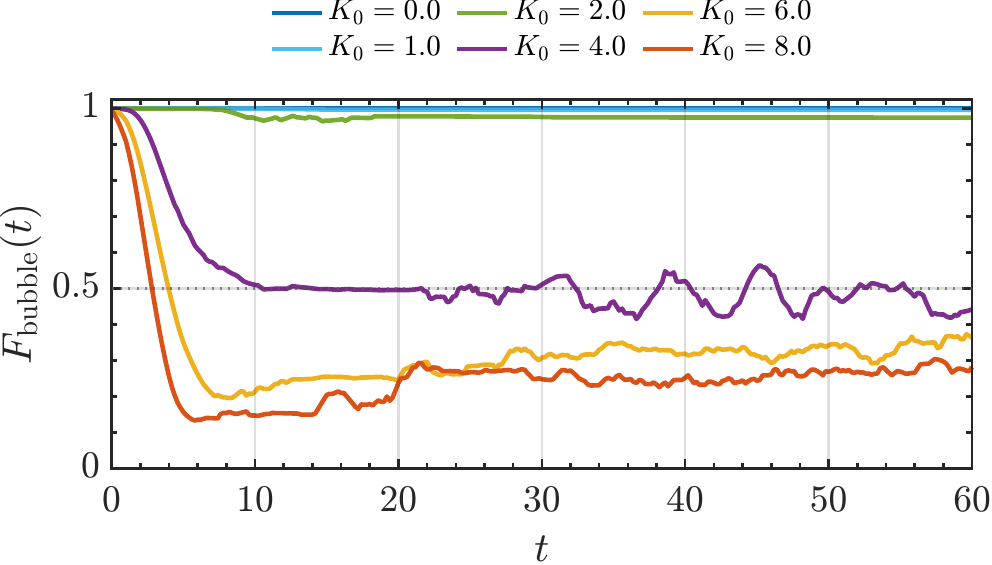}
\caption{Escape of part of the localized flat-top component under strong kicks. The
quantity \(F_{\mathrm{bubble}}(t)\), defined in
Eq.~\eqref{eq:FTS_remaining_fraction}, is the fraction of the conserved FTS norm
that remains within \(R=18\) of the tracked original bubble. A value of one
means that the FTS remains near the bubble, while smaller values indicate partial escape. For
\(K_0=0,1,\) and \(2\), which lie below the variational energetic estimate
\(K_{\mathrm{escape}}=3.0876\), the composite remains nearly intact.
For \(K_0=4,6,\) and \(8\), which lie above that scale, this fraction
drops sharply and subsequently exhibits inelastic fluctuations. Solid
curves are running means over \(\Delta t_{\mathrm{avg}}=2\), used only to
make the long-time response easier to read. The separation of the two
groups supports the variational prediction of an energetic onset of
substantial escape without implying a sharp dissociation threshold.}
\label{fig:remaining_fraction_time}
\end{figure}

Figure~\ref{fig:remaining_fraction_time} displays two clearly separated response
groups. Below the energetic scale, the late-time median fractions remaining near the bubble
for \(K_0=0,1,\) and \(2\) are \(1.000\), \(0.997\), and \(0.975\),
respectively. Above it, the corresponding values for \(K_0=4,6,\) and \(8\)
fall to \(0.423\), \(0.367\), and \(0.293\), and the remaining density shows
strong inelastic fluctuations. Thus, the simulations confirm the
qualitative change predicted by Eq.~\eqref{eq:K_energy_escape}: kicks below
the variational energetic scale leave the composite nearly intact, whereas
kicks above it cause a substantial part of the FTS to escape from the bubble.


\section{Conclusions}

We have introduced and studied a flat-top--bubble vector soliton in a pair
of repulsively coupled cubic-quintic fields. Although the coupling is
repulsive, it creates a bound composite structure. The localized flat-top
component pushes the background component away and opens a broad density
bubble. That bubble, in turn, provides a trapping region for the flat-top
component. The decoupled calculation supports this picture: when the
coupling is removed, the bubble disappears, while the localized component
remains as an isolated flat-top soliton. Direct real-time propagation also
shows that the stationary coupled state keeps its shape over long times.

We developed a background-subtracted variational description that remains
finite despite the nonzero background. Using normalized super-Gaussian
profiles, we reduced the translational dynamics to the motion of the two
component centers. Their overlap produces a binding well whose minimum is
the aligned state. The well is almost V-shaped over a broad range of
separations but has a smooth, parabolic bottom. A weak phase imprint applied
only to the localized component excites both center modes. The zero-frequency
Goldstone mode produces a common drift of the composite, whereas the
curvature of the locally parabolic binding well provides the restoring force
for the finite-frequency relative oscillation. The numerical simulations
confirm that the internal oscillation frequency changes very little across
the weak-kick range and agrees well with the variational prediction. An
independent energy minimization leads to the same overall picture.

We also followed the composite state beyond the small-oscillation regime.
Comparing the exact energy added by a phase imprint with the variational
binding energy gives a simple scale for the onset of partial
escape. Calculations of the fraction remaining near the bubble confirm this qualitative prediction:
below this scale, the localized flat-top component remains almost fully
bound to the bubble, whereas above it, a substantial part of its density
leaves the original bubble and the remaining state shows strong inelastic
fluctuations. The change is smooth and should therefore be viewed as the
onset of partial escape rather than a sharp breakup point. Taken together, these
results show that weak internal oscillations and the partial escape caused
by strong kicks are two limits of the same binding mechanism. They provide a
simple framework for studying self-induced vector structures and their
response to both weak and strong driving.


\appendix

\section{Modulational stability of the background}
\label{background_stability}

The finite background of the \(u\) component must be modulationally
stable. Setting \(v=0\) in Eq.~\eqref{eq:sub_explicit1} gives
\begin{equation}
i u_t
=
-\frac12 u_{xx}
-g\left(\abs{u}^2-\rho_0\right)u
+q\left(\abs{u}^4-\rho_0^2\right)u .
\label{eq:u_scalar_sub}
\end{equation}
The uniform solution is \(u=\sqrt{\rho_0}\). We perturb it according to
\begin{equation}
u(x,t)
=
\sqrt{\rho_0}+a(x,t)+i b(x,t),
\label{eq:linear_perturbation}
\end{equation}
where \(a\) and \(b\) are real and small. To linear order, Eq.~\eqref{eq:u_scalar_sub}
yields
\begin{equation}
a_t=-\frac12 b_{xx},
b_t=
\frac12 a_{xx}
-
2\rho_0\left(-g+2q\rho_0\right)a .
\label{eq:linear_ab}
\end{equation}
Using plane-wave perturbations proportional to
\(e^{i(Kx-\Omega t)}\), we obtain the dispersion relation
\begin{equation}
\Omega^2
=
\frac{K^4}{4}
+
K^2\rho_0\left(-g+2q\rho_0\right).
\label{eq:bogoliubov_dispersion}
\end{equation}
The uniform background is therefore modulationally stable provided
\begin{equation}
-g+2q\rho_0>0,
\;\;\text{or equivalently}\;\;
\rho_0>\frac{g}{2q}.
\label{eq:mi_condition}
\end{equation}
For the parameters used in the simulations, \(g=q=4\) and \(\rho_0=1.2\), this
condition is satisfied since
\begin{equation}
-g+2q\rho_0=5.6>0.
\end{equation}


\section{Constant term of the averaged Lagrangian}
\label{constant_term}

The averaged Lagrangian in Eq.~\eqref{eq:VA_L_full} contains the coordinate-independent term
\begin{align}
C_L
={}&
-\frac{N_1^2p}{8a_u^3\rho_0G_p^2}
\Gamma\left(2-\frac{1}{2p}\right)
\LerchPhi\left(z_u,s_p,2\right)
\nonumber\\
&+
\frac{2^{-2-1/(2p)}N_1^2\left(g-2q\rho_0\right)}{a_uG_p}
+
\frac{3^{-1-1/(2p)}qN_1^3}{4a_u^2G_p^2}
\nonumber\\
&-
\frac{mN_2}{4a_v^2G_m}
\Gamma\left(2-\frac{1}{2m}\right)
+
\frac{2^{-2-1/(2m)}gN_2^2}{a_vG_m}
\nonumber\\
&-
\frac{3^{-1-1/(2m)}qN_2^3}{4a_v^2G_m^2},
\label{eq:VA_CL}
\end{align}
where
\begin{equation}
s_p=2-\frac{1}{2p},
z_u=\frac{N_1}{2a_u\rho_0G_p},
\end{equation}
and \(\LerchPhi(z,s,a)\) is the Lerch transcendent
\cite{Erdelyi1953}.

\section{Static energetic minimization}
\label{app:static_minimization}

We adapt the direct variational energy-minimization procedure previously applied to stationary two-dimensional flat-top solitons \cite{AlotaibiPRE2026} to the present one-dimensional, two-component flat-top--bubble system. An independent estimate of the stationary profiles is then obtained by minimizing the background-subtracted energy over aligned one-dimensional super-Gaussian densities. The two densities are written as
\begin{align}
d_u(x)
&=
D_u
\exp\left[
-\left|\frac{x}{a_u}\right|^{2p}
\right],
\abs{u(x)}^2=\rho_0-d_u(x),
\label{eq:static_du_ansatz}
\\
\abs{v(x)}^2
&=
A_v^2
\exp\left[
-\left|\frac{x}{a_v}\right|^{2m}
\right],
A_v^2=\frac{N_2}{2a_vG_m}.
\label{eq:static_v_ansatz}
\end{align}
Here \(D_u\) is the depletion depth, \(a_u\) and \(a_v\) are the widths, and \(p\) and \(m\) are the shape parameters. They satisfy
\(0<D_u\leq\rho_0\), \(a_u,a_v>0\), and \(p,m>0\). The value
\(D_u=\rho_0\) represents a fully depleted bubble core. The deficit norm is
\begin{equation}
N_1=2D_u a_uG_p.
\label{eq:static_N1}
\end{equation}
For these static profiles, \(c_u=c_v=0\), so the background kinetic counterterm in Eq.~\eqref{eq:VA_Ldensity} vanishes.

The finite background-subtracted energy is
\begin{align}
E_{\mathrm{rel}}
={}&
\int_{-\infty}^{\infty}
\bigg[
\frac12\abs{u_x}^2
+
\frac12\abs{v_x}^2
\nonumber\\
&+
\frac12\left(-g+2q\rho_0\right)
\left(\abs{u}^2-\rho_0\right)^2
+
\frac{q}{3}
\left(\abs{u}^2-\rho_0\right)^3
\nonumber\\
&-
\frac{g}{2}\abs{v}^4
+
\frac{q}{3}\abs{v}^6
+
\kappa\abs{v}^2
\left(\abs{u}^2-\rho_0\right)
\bigg]\dd x .
\label{eq:static_Erel}
\end{align}
Substitution of Eqs.~\eqref{eq:static_du_ansatz} and
\eqref{eq:static_v_ansatz} gives
\begin{equation}
E_{\mathrm{var}}
=
E_{\mathrm{var}}(D_u,a_u,p,a_v,m).
\label{eq:static_Evar_function}
\end{equation}
This energy is separated into
\begin{equation}
E_{\mathrm{var}}
=
E_v
+
E_{u,\mathrm{grad}}
+
E_{u,\mathrm{pot}}
+
E_{\mathrm{int}}^{\mathrm{rel}}.
\label{eq:static_Evar}
\end{equation}
The localized-component contribution is
\begin{align}
E_v(a_v,m)
={}&
\frac{mN_2}{4a_v^2G_m}
\Gamma\left(2-\frac{1}{2m}\right)
\nonumber\\
&-
gA_v^4a_vG_m\,2^{-1/(2m)}
\nonumber\\
&+
\frac{2q}{3}A_v^6a_vG_m\,3^{-1/(2m)}.
\label{eq:static_Ev}
\end{align}
The potential energy of the depleted background is
\begin{align}
E_{u,\mathrm{pot}}(D_u,a_u,p)
={}&
\left(-g+2q\rho_0\right)
D_u^2a_uG_p\,2^{-1/(2p)}
\nonumber\\
&-
\frac{2q}{3}
D_u^3a_uG_p\,3^{-1/(2p)}.
\label{eq:static_Eupot}
\end{align}
The background gradient energy is
\begin{equation}
E_{u,\mathrm{grad}}
=
\frac18
\int_{-\infty}^{\infty}
\frac{\left[d_u'(x)\right]^2}
{\rho_0-d_u(x)}
\dd x,
\label{eq:static_Eugrad}
\end{equation}
where
\(\abs{u_x}^2=[d_u'(x)]^2/[4(\rho_0-d_u)]\).
At \(D_u=\rho_0\), the integrand is defined by its continuous limit at
\(x=0\), and the integral remains finite for \(p>1/2\).

The interaction energy is
\begin{align}
E_{\mathrm{int}}^{\mathrm{rel}}
={}&
-\kappa D_uA_v^2
\int_{-\infty}^{\infty}
\exp\left[-\left|\frac{x}{a_v}\right|^{2m}\right]
\nonumber\\
&\times
\exp\left[-\left|\frac{x}{a_u}\right|^{2p}\right]
\dd x.
\label{eq:static_Eint}
\end{align}
This integral is evaluated numerically when \(m\neq p\).

We vary the five parameters within the ranges stated above and select the values that give the lowest energy:
\begin{equation}
E_{\mathrm{var}}^{\min}
=
E_{\mathrm{var}}
\left(D_u^{*},a_u^{*},p^{*},a_v^{*},m^{*}\right),
\label{eq:static_argmin}
\end{equation}
where the superscript \(*\) denotes the minimizing values. 

If all five minimizing values lie within their allowed ranges, they satisfy
\[
\frac{\partial E_{\mathrm{var}}}{\partial D_u}
=
\frac{\partial E_{\mathrm{var}}}{\partial a_u}
=
\frac{\partial E_{\mathrm{var}}}{\partial p}
=
\frac{\partial E_{\mathrm{var}}}{\partial a_v}
=
\frac{\partial E_{\mathrm{var}}}{\partial m}
=0.
\]

If the minimum occurs at \(D_u=\rho_0\), the depletion depth is fixed by the boundary rather than by
\(\partial E_{\mathrm{var}}/\partial D_u=0\). Terms without a closed analytic form are evaluated by numerical integration.

For
\begin{equation}
g=q=4,
\rho_0=1.2,
\kappa=4,
N_2=30,
\label{eq:static_parameters}
\end{equation}
the minimizing values are compared with the imaginary-time profile fit in Table~\ref{tab:static_min_results}. The minimum occurs at
\(D_u=\rho_0\), corresponding to a fully depleted bubble core.

\begin{table}[t]
\centering
\caption{Parameters obtained from the imaginary-time profile fit and the energy minimization.}
\begin{tabular}{lccccc}
\hline
method & \(D_u\) & \(a_u\) & \(p\) & \(a_v\) & \(m\) \\
\hline
profile fit
& \(1.2000\)
& \(12.9298\)
& \(22.7550\)
& \(12.6487\)
& \(21.5548\)
\\
energy minimum
& \(1.2000\)
& \(12.8927\)
& \(24.7270\)
& \(12.6612\)
& \(17.4711\)
\\
\hline
\end{tabular}
\label{tab:static_min_results}
\end{table}

Equation~\eqref{eq:static_N1} gives
\begin{equation}
N_1^{(E)}=30.59365>N_2
\label{eq:static_min_N1}
\end{equation}
for the energy-minimized state, compared with \(N_1=30.65238\) for the imaginary-time state. The corresponding energies are
\(E_{\mathrm{var}}^{\min}=-112.0230\),
\(-111.9542\) for the same functional evaluated at the profile-fit parameters, and
\(E_{\mathrm{ITP}}=-112.1435\) for the imaginary-time state.

For a super-Gaussian density with width \(a\) and order \(r\), the full width at half maximum is
\begin{equation}
W_{\mathrm{FWHM}}
=
2a(\ln 2)^{1/(2r)}.
\label{eq:static_FWHM}
\end{equation}
The energy-minimized widths are
\(W_{u,\mathrm{FWHM}}=25.60\) and
\(W_{v,\mathrm{FWHM}}=25.06\), compared with the imaginary-time values
\(25.5\) and \(25.0\), respectively. 

Finally, inserting the energy-minimized parameters and \(N_1^{(E)}\) into Eq.~\eqref{eq:VA_general_frequency} gives
\begin{equation}
\omega_{\mathrm{VA}}^{(E)}=0.07296,
T_{\mathrm{VA}}^{(E)}=86.12.
\label{eq:static_min_frequency}
\end{equation}
The difference from
\(\omega_{\mathrm{num}}=0.07948\) is \(8.20\%\), compared with
\(4.99\%\) for the profile-fit prediction in
Eq.~\eqref{eq:VA_numerical_prediction}. The energy minimization therefore provides a test that is independent of the profile-fitting procedure, although its frequency prediction is slightly less accurate.



\begin{thebibliography}{99}

\bibitem{Manakov1974}
S. V. Manakov,
``On the theory of two-dimensional stationary self-focusing of electromagnetic waves,''
Sov. Phys. JETP \textbf{38}, 248--253 (1974).

\bibitem{ChristodoulidesJoseph1988}
D. N. Christodoulides and R. I. Joseph,
``Vector solitons in birefringent nonlinear dispersive media,''
Opt. Lett. \textbf{13}, 53--55 (1988).

\bibitem{KivsharLutherDavies1998}
Y. S. Kivshar and B. Luther-Davies,
``Dark optical solitons: Physics and applications,''
Phys. Rep. \textbf{298}, 81--197 (1998).

\bibitem{Frantzeskakis2010}
D. J. Frantzeskakis,
``Dark solitons in atomic Bose--Einstein condensates: From theory to experiments,''
J. Phys. A: Math. Theor. \textbf{43}, 213001 (2010).

\bibitem{KevrekidisBook}
P. G. Kevrekidis, D. J. Frantzeskakis, and R. Carretero-Gonzalez,
\textit{The Defocusing Nonlinear Schr\"odinger Equation: From Dark Solitons to Vortices and Vortex Rings}
(SIAM, Philadelphia, 2015).

\bibitem{BuschAnglin2001}
Th. Busch and J. R. Anglin,
``Dark--bright solitons in inhomogeneous Bose--Einstein condensates,''
Phys. Rev. Lett. \textbf{87}, 010401 (2001).

\bibitem{Becker2008}
C. Becker, S. Stellmer, P. Soltan-Panahi, S. D\"orscher,
M. Baumert, E.-M. Richter, J. Kronj\"ager, K. Bongs, and K. Sengstock,
``Oscillations and interactions of dark and dark--bright solitons in Bose--Einstein condensates,''
Nat. Phys. \textbf{4}, 496--501 (2008).

\bibitem{AlotaibiCarr2017}
M. O. D. Alotaibi and L. D. Carr,
``Dynamics of dark-bright vector solitons in Bose--Einstein condensates,''
Phys. Rev. A \textbf{96}, 013601 (2017).

\bibitem{Malomed2022}
B. A. Malomed,
``Soliton models: Traditional and novel, one- and multidimensional,''
Low Temp. Phys. \textbf{48}, 856--895 (2022).

\bibitem{BirnbaumMalomed2008}
Z. Birnbaum and B. A. Malomed,
``Families of spatial solitons in a two-channel waveguide with the cubic-quintic nonlinearity,''
Physica D \textbf{237}, 3252--3262 (2008).

\bibitem{AlotaibiPRE2026}
M. O. D. Alotaibi, Y. O. A. Abughnheim, L. Al Sakkaf, and U. Al Khawaja,
``Phase-controlled elastic, inelastic, and coalescent collisions of two-dimensional flat-top solitons,''
Phys. Rev. E \textbf{113}, 054202 (2026).

\bibitem{Petrov2015}
D. S. Petrov,
``Quantum mechanical stabilization of a collapsing Bose-Bose mixture,''
Phys. Rev. Lett. \textbf{115}, 155302 (2015).

\bibitem{Katsimiga2023}
G. C. Katsimiga, S. I. Mistakidis, B. A. Malomed,
D. J. Frantzeskakis, R. Carretero-Gonzalez, and P. G. Kevrekidis,
``Interactions and dynamics of one-dimensional droplets, bubbles and kinks,''
Condens. Matter \textbf{8}, 67 (2023).

\bibitem{Tylutki2020}
M. Tylutki, G. E. Astrakharchik, B. A. Malomed, and D. S. Petrov,
``Collective excitations of a one-dimensional quantum droplet,''
Phys. Rev. A \textbf{101}, 051601(R) (2020).

\bibitem{ZengFractional2021}
L. Zeng, B. A. Malomed, D. Mihalache, Y. Cai,
X. Lu, Q. Zhu, and J. Li,
``Bubbles and W-shaped solitons in Kerr media with fractional diffraction,''
Nonlinear Dyn. \textbf{104}, 4253--4264 (2021).

\bibitem{ZengFlatFloor2021}
L. Zeng, B. A. Malomed, D. Mihalache, Y. Cai,
X. Lu, Q. Zhu, and J. Li,
``Flat-floor bubbles, dark solitons, and vortices stabilized by inhomogeneous nonlinear media,''
Nonlinear Dyn. \textbf{106}, 815--830 (2021).

\bibitem{FilatrellaMalomedSalerno2014}
G. Filatrella, B. A. Malomed, and M. Salerno,
``Domain walls and bubble droplets in immiscible binary Bose gases,''
Phys. Rev. A \textbf{90}, 043629 (2014).

\bibitem{Paredes2025}
A. Paredes, J. Guerra-Carmenate, J. R. Salgueiro,
D. Tommasini, and H. Michinel,
``Traveling bubbles and vortex pairs within symmetric two-dimensional quantum droplets,''
Phys. Rev. E \textbf{111}, 054217 (2025).

\bibitem{ParedesBoundary2025}
A. Paredes, J. Guerra-Carmenate, and H. Michinel,
``Fate of traveling waves at the boundary of quantum droplets,''
Phys. Rev. E \textbf{112}, 054205 (2025).

\bibitem{KivsharKrolikowski1995}
Y. S. Kivshar and W. Kr\'olikowski,
``Lagrangian approach for dark solitons,''
Opt. Commun. \textbf{114}, 353--362 (1995).

\bibitem{Erdelyi1953}
A. Erd\'elyi, W. Magnus, F. Oberhettinger, and F. G. Tricomi,
\textit{Higher Transcendental Functions}, Vol.~I
(McGraw--Hill, New York, 1953), Sec.~1.11.

\end{thebibliography}
\end{document}